# Application of Structural Similarity Analysis of Visually Salient Areas and Hierarchical Clustering in the Screening of Similar Wireless Capsule Endoscopic Images

Rui Nie[2‡]，Huan Yang[1‡]，Hejuan Peng[2]，Wenbin Luo[2]，Weiya Fan[2]，Jie Zhang[2]，Jing Liao[2]，Fang Huang[2*]，Yufeng Xiao[1*]

[‡]Both authors contribute equally to this work.

[1]Department of Gastroenterology，Second Affiliated Hospital，Army Medical University（Third Military Medical University），Chongqing，400037.

[2]Chongqing Jinshan Science & Technology (Group) Co., Ltd.，Chongqing，401120.

[*]Corresponding authors：Fang Huang，e-mail:modinaxi@163.com

Yufeng Xiao，e-mail:jackshawxyf@163.com

**Abstract**: Small intestinal capsule endoscopy is the mainstream method for inspecting small intestinal lesions, but a single small intestinal capsule endoscopy will produce 60,000–120,000 images, the majority of which are similar and have no diagnostic value. It takes 2–3 hours for doctors to identify lesions from these images. This is time-consuming and increases the probability of misdiagnosis and missed diagnosis since doctors are likely to experience visual fatigue while focusing on a large number of similar images for an extended period of time. In order to solve these problems, we proposed a similar wireless capsule endoscope (WCE) image screening method based on structural similarity analysis and the hierarchical clustering of visually salient sub-image blocks. The similarity clustering of images was automatically identified by hierarchical clustering based on the hue, saturation, value (HSV) spatial color characteristics of the images, and the keyframe images were extracted based on the structural similarity of the visually salient sub-image blocks, in order to accurately identify and screen out similar small intestinal capsule endoscopic images. Subsequently, the proposed method was applied to the capsule endoscope imaging workstation. After screening out similar images in the complete data gathered by the Type I OMOM Small Intestinal Capsule Endoscope from 52 cases covering 17 common types of small intestinal lesions, we obtained a lesion recall of 100% and an average similar image reduction ratio of 76%. With similar images screened out, the average play time of the OMOM image workstation was 18 minutes, which greatly reduced the time spent by doctors viewing the images.

# 1. INTRODUCTION

Moving with the peristalsis of the intestine, the small intestinal capsule endoscope captures images at a fixed or variable frame rate. It takes approximately 10 hours for the endoscope to capture all the images needed. Each case involves about 60,000–120,000 small intestinal capsule endoscopic images, and it takes an average of 2–3 hours for a doctor to diagnose a patient, which means that the doctor suffers from a huge workload when viewing these images. High labor and time costs hinder the popularization of medical small intestinal capsule endoscopes. Due to the continuous development of technology, and in order to capture the images more comprehensively and reduce missed shots, medical small intestinal capsule endoscopes have achieved a higher frame rate and image resolution, which means the number of images per patient will increase exponentially and the doctors' workload when viewing the images will increase. Decreasing the number of images viewed by doctors while increasing the detection rate of lesions has become an important focus of research in this field. Since the captured images are spatiotemporally continuous, a large number of adjacent images are highly similar. If similar small intestinal capsule endoscopic images could be accurately identified and screened out and the integrity of the small intestinal mucosa could be maintained without missing images of abnormal lesions, the doctors' workload would be dramatically reduced.

Many studies have been carried out on the screening of similar small intestinal capsule endoscopic images. Zhou et al. [1, 2] used keyframe extraction to remove redundant images. Tsevas et al. [3–5] applied the optical flow method to estimate the motion of the capsule and identify the backward redundant images among the capsule endoscopic images. Alexandros [6] used the Bayer image to reduce the dimensions of the original data and evaluated image similarity using local binary patterns (LBP) and the color structure similarity measurement method. Lee et al. [7] proposed a similarity confidence function based on area segmentation and global mapping to identify the registration area of WCE images, so as to identify the similarity of adjacent image data. Lee et al. [8] proposed a WCE motion model to estimate the movement of continuous WCE images, identify the key images in the video, and retain scene changes, thereby reducing the number of images. These research efforts have yielded results in identifying the similarity of small intestinal capsule endoscopic images.

Another research approach is to decrease the workload of doctors viewing WCE images by identifying abnormal lesions. Liu et al. [9–13] applied the currently popular convolutional neural network technology to classify and identify lesions in small intestinal capsule endoscopic images. The uninterpretable and data-driven nature of deep neural networks, however, often results in greatly reduced recognition accuracy in practical application scenarios. Hence, many problems urgently need to be solved before applying deep learning technology to the clinical practice of small intestinal capsule endoscopy. Zhou et al. [14, 15] applied traditional machine learning as opposed to the deep neural networks in order to identify abnormalities in small intestinal capsule endoscopy.

The methods based on similarity recognition or lesion recognition are currently confronted with the practical problem of designing and verifying a single lesion or several lesions. Doctors must diagnose a wide range of small intestinal lesions, however, so the optimal image screening scheme for doctors should preserve intestinal mucosal integrity as much as possible without missing any lesions.

In this study, the hierarchical clustering method [16, 17] was used to cluster the small intestinal capsule endoscopic images, and visual saliency model theory [18–20] was introduced to identify the visually salient areas in small intestinal capsule endoscopic images. Based on the structural similarity index (SSIM) principle [21], a small intestinal capsule endoscopic keyframe image extraction method was proposed based on hierarchical clustering and the structural similarity of visually salient sub-image blocks, reserving continuous small intestinal capsule endoscopic images for doctors to review without missing images with lesions. Section 2 introduces the procedure for clustering similar images based on hierarchical clustering, and the method of extracting small intestinal capsule endoscopic keyframe images based on the structural similarity of visually salient sub-image blocks. In Section 3, the test dataset of 17 common small intestinal lesions is established. In this study, subjective and objective evaluation indices based on the test set were designed to comprehensively evaluate the proposed algorithm, which was then applied to screen out redundant images from the complete dataset of 52 cases gathered by the Type I OMOM Small Intestinal Capsule Endoscope produced by Chongqing Jinshan Science & Technology (Group) Co., Ltd. in order to verify the effectiveness of the algorithm. The experimental results are discussed in Section 4. Section 5 is a summary of the study.

## 2. ALGORITHMIC THEORY

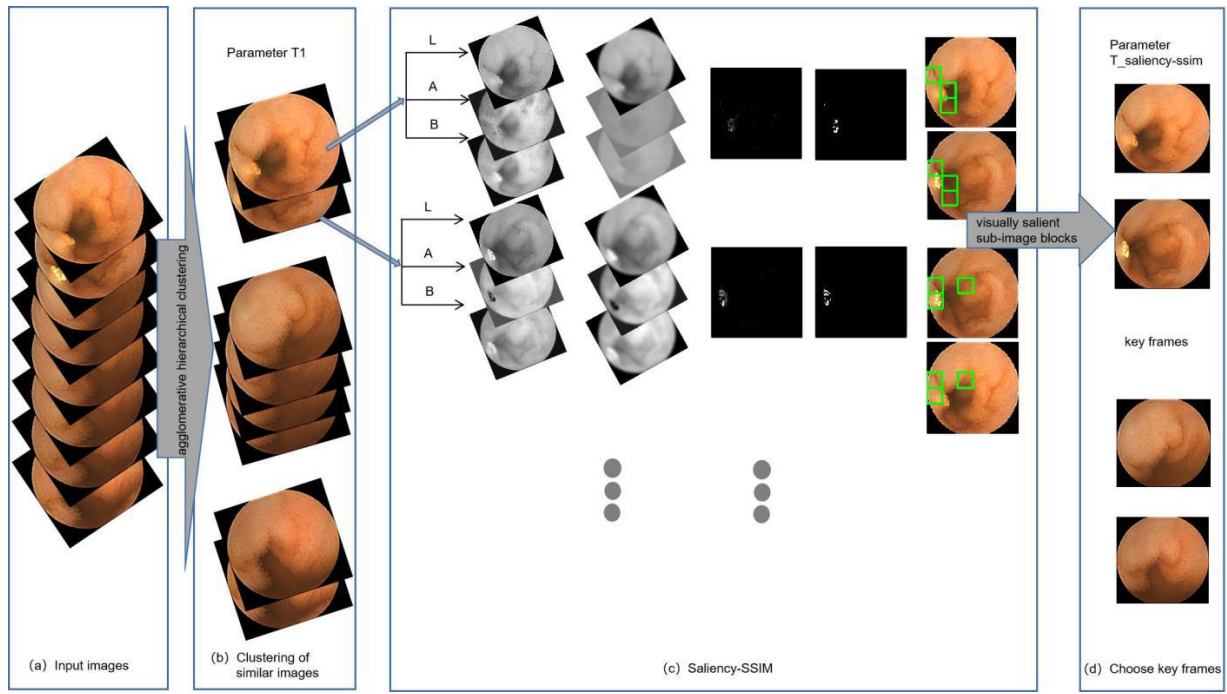

**Figure 1. Block diagram of the algorithm**

Figure 1 presents a block diagram of the algorithm proposed in this study. First, the color features of the entire images were extracted in order to cluster similar images based on hierarchical clustering. Then, the visually salient sub-image blocks were extracted from images in the cluster. The structural similarities between the sub-image blocks of adjacent images were compared to accurately identify keyframe images and screen out non-keyframe images.

### 2.1 Clustering of similar small intestinal capsule endoscopic images based on agglomerative hierarchical clustering

#### *2.1.1 Extraction of color features*

The color space of the images was converted to hue, saturation, value (HSV) space, where the histogram features of the 3 channels were extracted, the value of each channel was quantized into 10 intervals in order to count the number in each interval, the value of each interval was normalized, and a 1*1000 color feature vector q was ultimately extracted.

### 2.1.2 Clustering of similar images

Every n images out of the N small intestinal capsule endoscopic images, $\{x_1, x_2, \ldots, x_N\}$, were divided in chronological order into a cluster denoted as $D\{x_1, x_2, \ldots, x_n\}$. The images in D were converted to the HSV color space and the color feature vectors, $Q\{q_1, q_2, \ldots, q_n\}$, were extracted. The color feature space distance tree was constructed as follows: First, the n images in cluster D were regarded as n categories, the L2 distance between the color feature vectors was calculated, and the 2 images with the smallest distance were merged into the same category, with each image regarded as a leaf node of the distance tree. The distance between the merged category and all the remaining categories was then recalculated, and the above steps were repeated until all the images had been merged into 1 category. After the completion of the above steps, the distance tree was constructed. The maximum distance of the distance tree was denoted as D, T1 represents the threshold, and images with a distance < T1×D in the tree were classified into the cluster of similar images.

## 2.2 Saliency-SSIM theory

First, image $img_i$ was converted to the Lab color space in order to obtain 3 channel components, i.e., L(x,y), A(x,y), and B(x,y). Gaussian filtering was then performed on the 3 channels:

$$G(x, y) = \frac{1}{2\pi\sigma^2} \exp(-\frac{x^2 + y^2}{2\sigma^2}) \quad (1)$$

The 3 filtered channels are L_g(x,y), A_g(x,y), and B_g(x,y).

On the basis of the saliency mathematical model established by R. Achanta et al. [20] based on the Lab color model, the following equation was obtained:

$$S(x,y) = \| I_u - I_{Lab}(x,y) \| \qquad (2)$$

We calculated the significance of the Lab color model for L_g(x,y), A_g(x,y), and B_g(x,y), respectively:

$$S_L(x,y) = \| L(x,y) - L\_g(x,y) \| \qquad (3)$$

$$S_A(x,y) = \| A(x,y) - A\_g(x,y) \| \qquad (4)$$

$$S_B(x,y) = \| B(x,y) - B\_g(x,y) \| \qquad (5)$$

The values of the 3 channels, i.e., $S_L(x,y)$, $S_A(x,y)$, and $S_B(x,y)$, were combined to obtain the image edge information Img_g. The sum of the mean and standard deviation of Img_g was calculated and recorded as Threshold. With Threshold as the threshold, Img_g was binarized to obtain the visually salient pixels in the image. Img_g was divided into 40 × 40-pixel sub-image blocks denoted Img_g1, Img_g2…Img_g36. By counting the number of significant pixels in each sub-image block, the regional coordinates of the 3 sub-image blocks containing the largest number of the most significant pixels were extracted, i.e., area1(x1,y1,x2,y2), area2(x1,y1,x2,y2), and area3(x1,y1,x2,y2). Based on the regional coordinates of the sub-image blocks, the visually salient sub-image blocks of the original images $img_i$ and $img_{i+1}$ were extracted and denoted as $s1_i(x,y)$, $s2_i(x,y)$, $s3_i(x,y)$ and $s1_{i+1}(x,y)$, $s2_{i+1}(x,y)$, $s3_{i+1}(x,y)$, respectively (Figure 2).

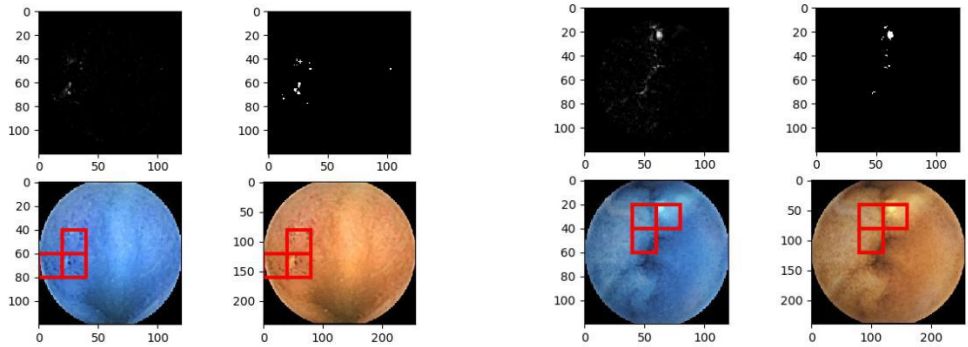

**Figure 2. Visually salient sub-image blocks**

Based on the MSSIM algorithm [21], we evaluated the similarity value T between the current image of the same cluster and the visually salient sub-image block of the next adjacent image. If T was greater than the threshold $T_{\_saliency\text{-}ssim}$, it would be identified as a keyframe in the cluster.

## 3. RESULTS

### 3.1 Small intestinal capsule endoscopic image test dataset

In this study, a small intestinal capsule endoscopic image test dataset including various types of small intestinal lesions (hereafter referred to as "the test set") was established. Among the complete data of 52 cases gathered by Chongqing Xin Qiao Hospital from January–December 2019 using a Type I OMOM Small Intestinal Capsule Endoscope produced by Chongqing Jinshan Science & Technology (Group) Co., Ltd., the doctors labeled all images with lesions, marked the sequence numbers of those images, and intercepted a total of 68,400 images in the lesion and the normal continuous video segment to form the test set. The test set comprised common types of small intestinal lesions, i.e., venous abnormalities, nodules, lumps, polyps, erythema, spots, white spots, erosions, ulcers, aphthous ulcers, diverticula, granules, redness, abnormal villi, paleness, hookworms, and bleeding (classified according to the CEST classification standard [22]). Figure 3 presents images of various types of small intestinal lesions. Table 1 lists the distribution of lesions in the test set.

**Table 1. Distribution of 52 WCE lesions**

| CEST | patient(s) | images |
|---|---|---|
| Protruding lesions-venous structure | 7 | 2422 |
| Protruding lesions-Nodule | 37 | 4646 |
| Protruding lesions-Mass/tumor | 14 | 626 |
| Protruding lesions-Polyp(s) | 3 | 158 |
| Flat lesions-Plaque(red) | 8 | 350 |
| Flat lesions-Spot | 5 | 541 |
| Flat lesions-Plaque(white) | 13 | 279 |
| Excavated lesion-Erosion | 11 | 872 |
| Excavated lesion-Ulcer | 7 | 217 |
| Excavated lesion-Aphtha | 5 | 369 |
| Excavated lesion-Diverticulum | 1 | 72 |
| Mucosa-Granular | 23 | 1048 |
| Mucosa-Erythematous | 5 | 1826 |
| Mucosa-Edematous(congested) | 4 | 212 |
| Mucosa-Pale | 2 | 38 |
| Content-Parasites | 1 | 33 |
| Content-Blood | 2 | 1067 |

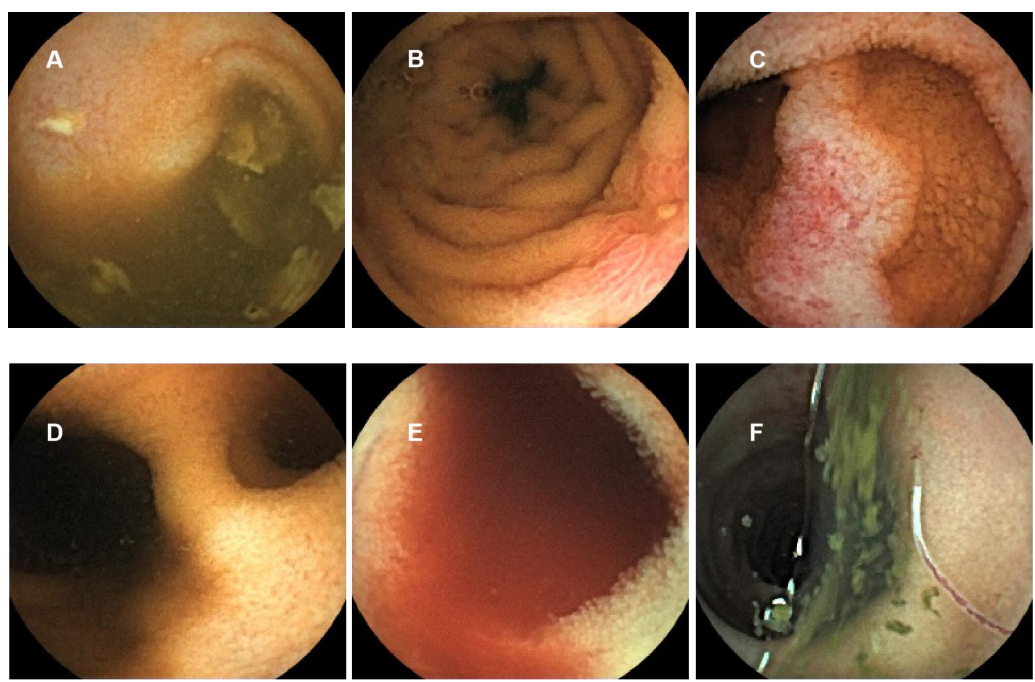

A:Excavated lesion-Aphtha
B:Excavated lesion-Ulcer
C:Excavated lesion-Erosion
D:Excavated lesion-Diverticulum
E:Content-Blood
F:Content-Parasites

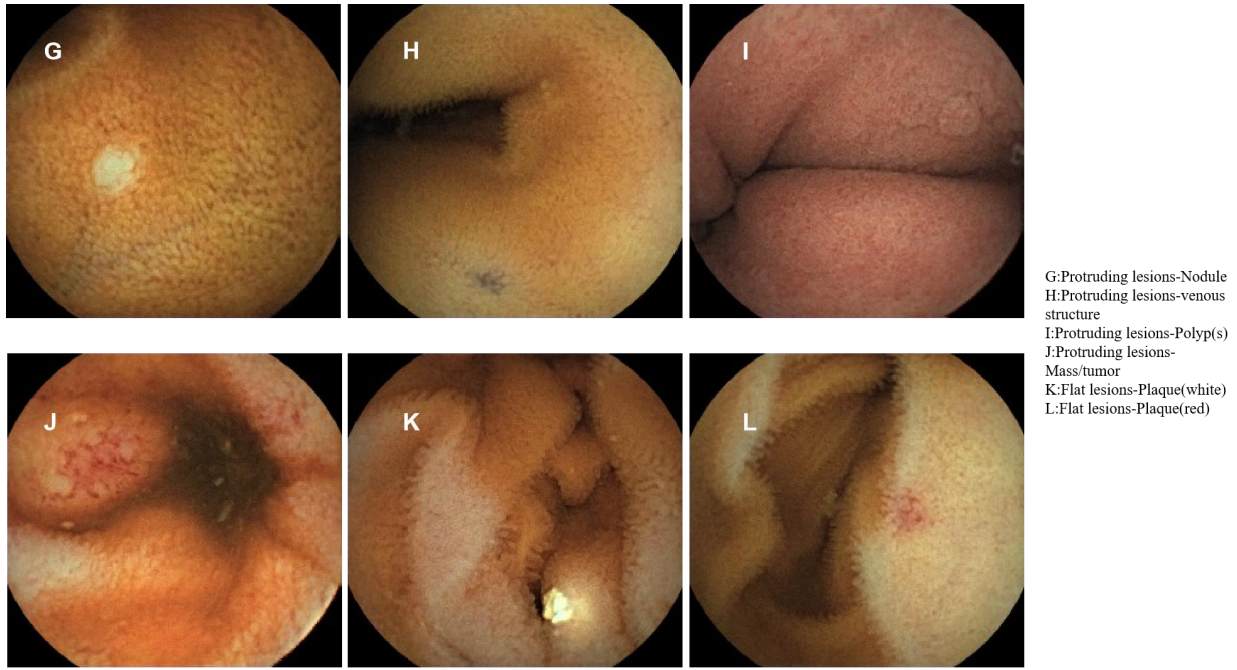

Figure 3. CEST classification of the test set

### 3.2 Evaluation indices

The similarity between images can be evaluated in 2 ways, i.e., subjective evaluation and algorithmic evaluation. Correspondingly, the image reduction ratio based on the principle of image similarity can also be divided into the subjective reduction ratio and the algorithmic image reduction ratio. The subjective reduction ratio was based on the gastroenterologists' subjective judgment of image similarity. The similar images in the same section of the test set were grouped together by 5 doctors independently, and it was ensured that no lesion was missed after clustering. The average of the subjective reduction ratios created by the 5 doctors was used as the subjective reduction ratio of the test set in this study. The subjective reduction ratio (*subjective_reduction_ratio*) was defined and calculated as follows:

$$subjective\_reduction\_ratio = 1 - \sum \frac{N_{cluster}}{N_{total}}, \quad (6)$$

where $N_{cluster}$ is the number of clusters of the test set selected by a single doctor, and $N_{total}$ is the total number of images in the test set.

Correspondingly, the reduction ratio of the algorithm for the test set was recorded as *ER_rate*, which was calculated as follows:

$$ER\_rate = 1 - \sum \frac{N_{key}}{N_{total}}, \quad (7)$$

where $N_{key}$ is the number of keyframe images in the test set selected by this algorithm.

Each image in the test set was represented by a unique sequence number. The doctors selected the images with lesions in the test set and classified the images per the capsule endoscopy structured terminology (CEST) classification standard [22]. $Ab_i(im_1, im_2……im_n)$ denotes the set of sequence numbers of images with the same type of lesions from the same case, $Ab(Ab_1, Ab_2, Ab_3……Ab_k)$ represents the set of lesions in the test set, $K$ is the number of elements in the lesion set $Ab$, $S(s1, s2, s3…… sq)$ denotes the set of sequence numbers of the keyframe images selected by the algorithm, and *SD* denotes the number of lesion categories included in the set *S*:

$$SD = \sum_{j=1}^{j=k} A_j,$$

$$A_j = \begin{cases} 1 & if\ Ab_j \bigcap S \neq \phi \\ 0 & else \end{cases} \quad (8)$$

The abnormal recall was defined as:

$$abnormal\_recall = \frac{SD}{K} \quad (9)$$

*T1* and *T_saliency-ssim* are parameters of the algorithm.

The HSV color features of the images in the test set were extracted in order to cluster the image data using the agglomerative hierarchical clustering method, the similarity between the images was evaluated using the clustering threshold $T1$, and the images were divided into different clusters. Saliency-SSIM keyframe image extraction was performed based on clustering. With threshold $T\_{saliency\text{-}ssim}$ as the reference, the keyframe images in the test set were identified, and indices such as *abnormal_recall* and *ER_rate* were calculated. Figure 4 shows the variation curves of *abnormal_recall* with *ER_rate* values having different parameters; $T1$ and $T\_{saliency\text{-}ssim}$. $T1$ ranged from 0.18–0.88, and $T\_{saliency\text{-}ssim}$ ranged from 0.03–0.12.

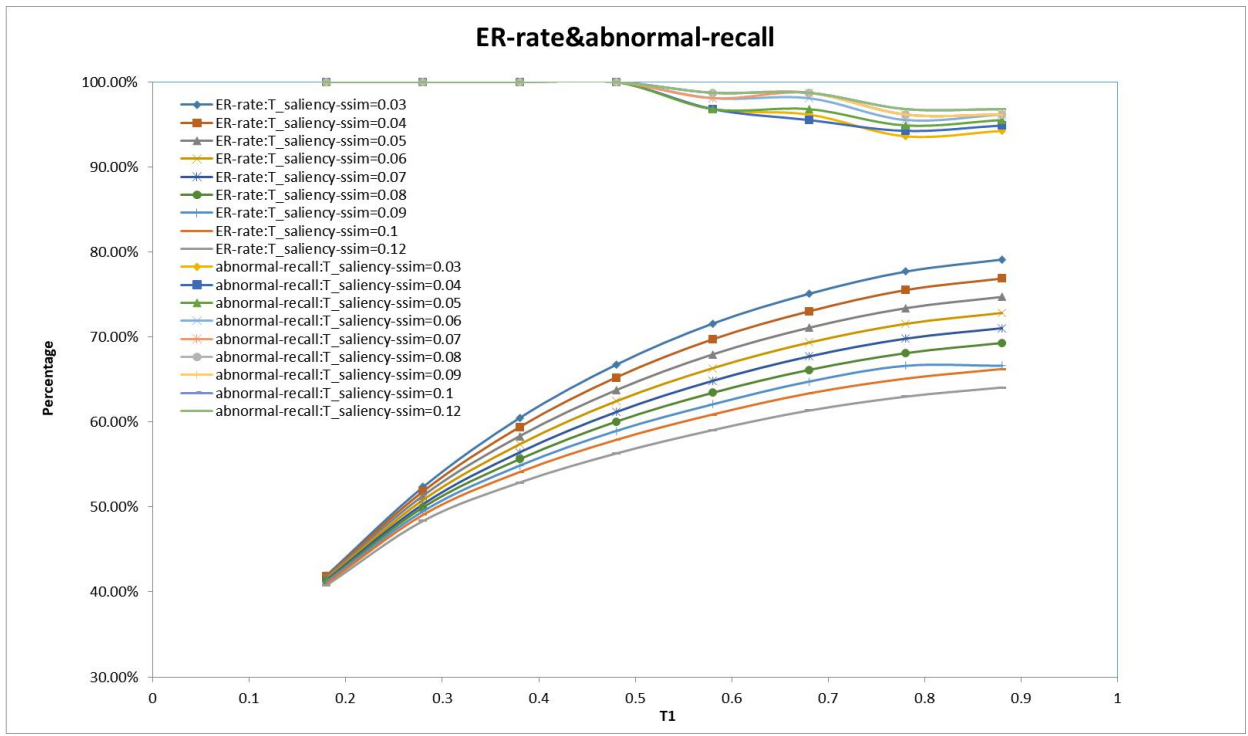

**Figure 4. Variation curves of image reduction ratio and abnormal recall with different algorithm parameters**

## Protruding lesions-Nodule

## Protruding lesions-Nodule

## Protruding lesions-venous structure

## Protruding lesions-venous structure

## Protruding lesions-Mass/tumor

## Protruding lesions-Mass/tumor

## Protruding lesions-Polyp(s)

## Protruding lesions-Polyp(s)

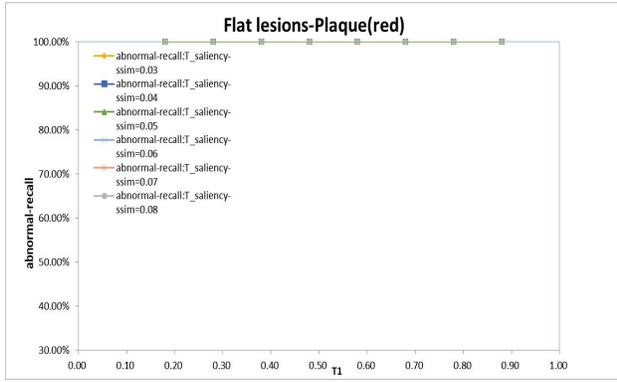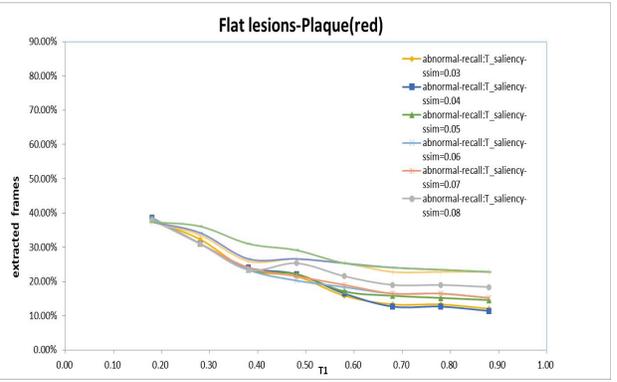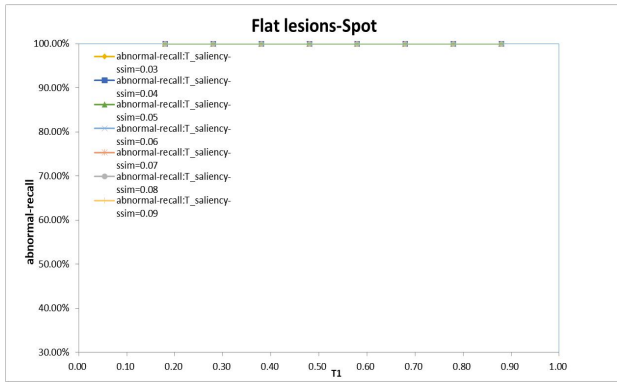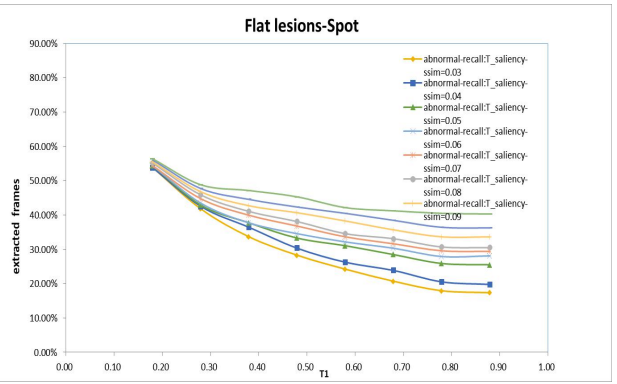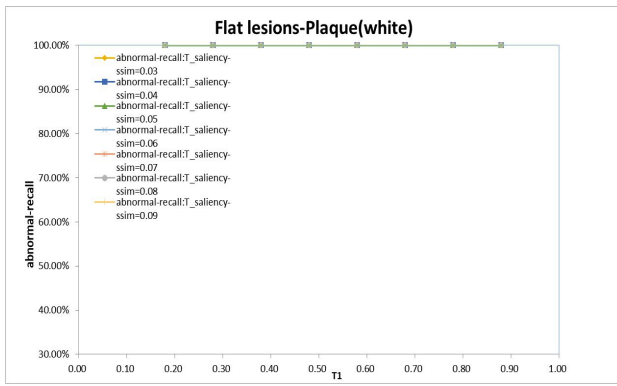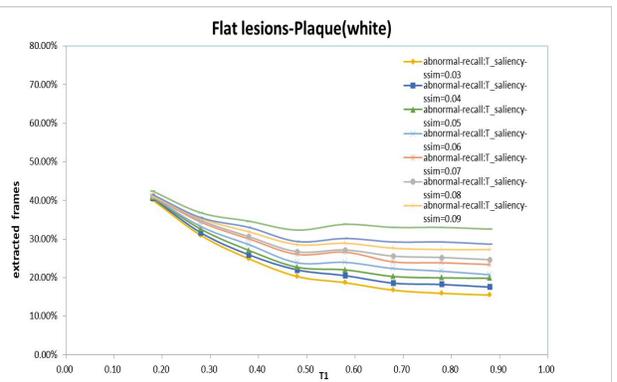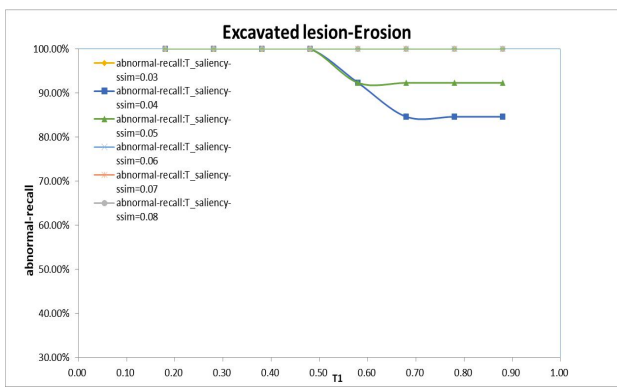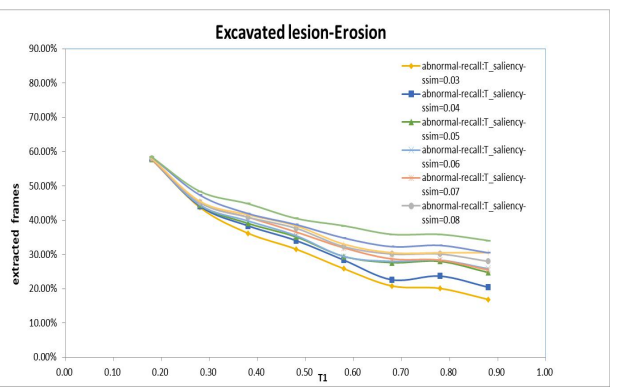

## Excavated lesion-Ulcer

## Excavated lesion-Ulcer

## Excavated lesion-Aphtha

## Excavated lesion-Aphtha

## Excavated lesion-Diverticulum

## Excavated lesion-Diverticulum

## Mucosa-Granular

## Mucosa-Granular

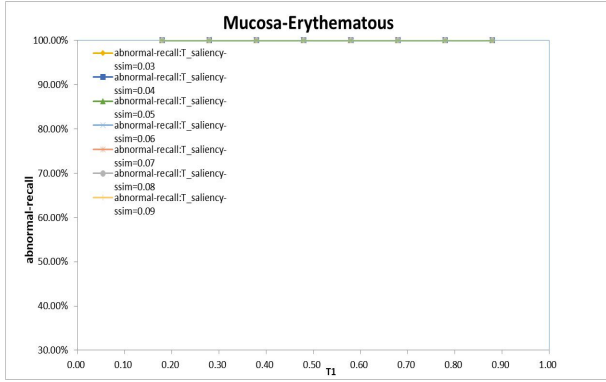
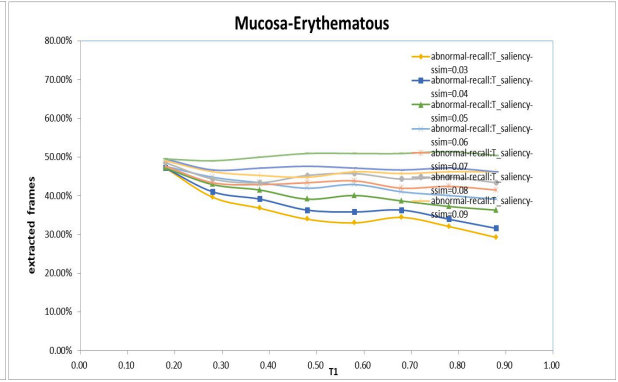
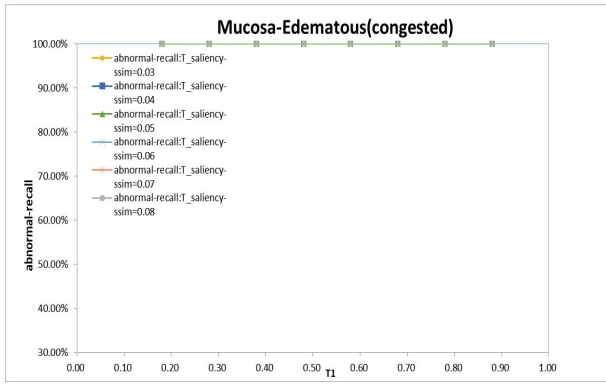
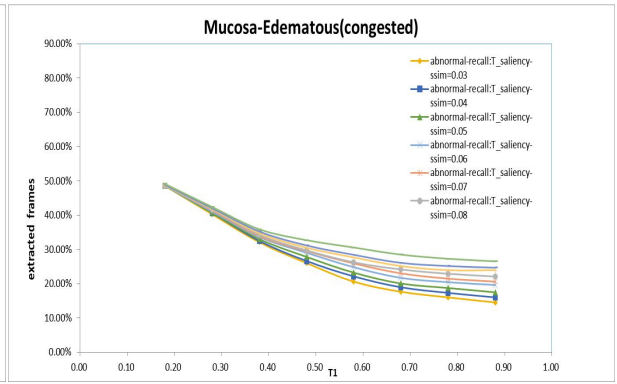
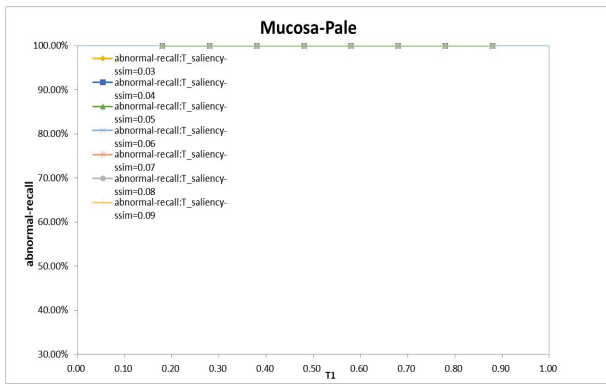
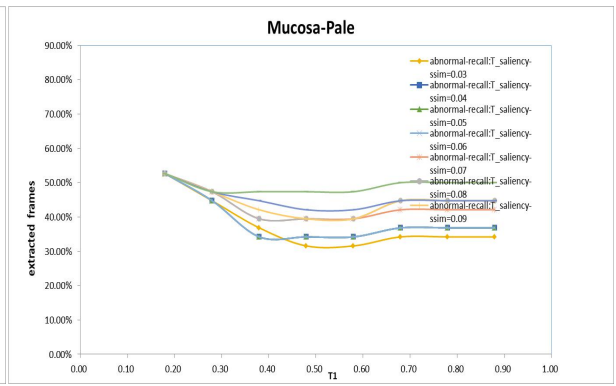
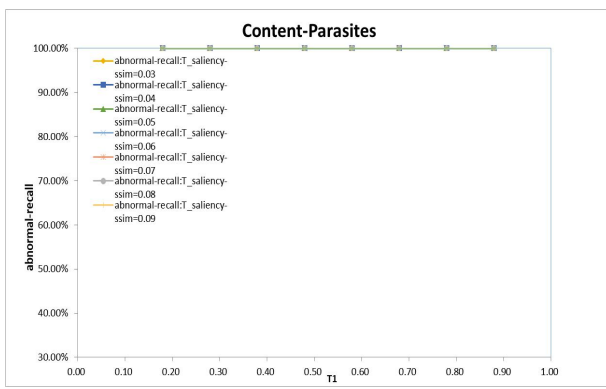
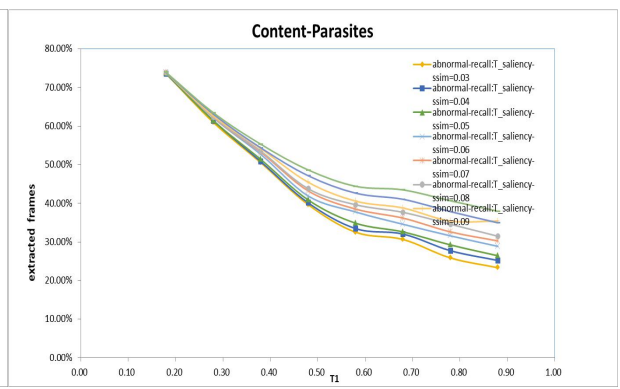

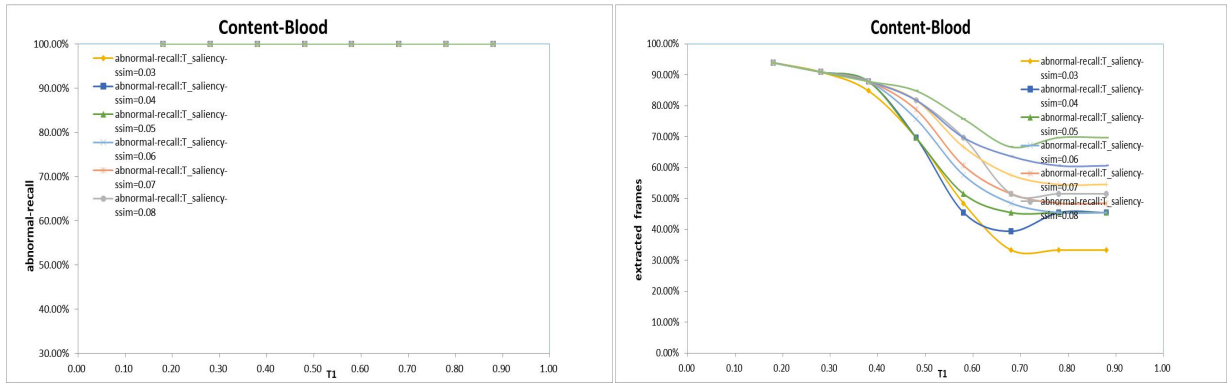

**Figure 5. Variation curves of image reduction ratio and abnormal recall**

Figure 5 illustrates the variation curves of the abnormal recall and image reduction ratios for all lesions in the test set, as well as the variation curves of the abnormal recall and image reduction ratios for a single lesion.

### 3.3 Results

As for the subjective evaluation conducted in this study, the 5 doctors independently divided the test set into 14,390, 13,556, 13,376, 12,987, and 14,672 clusters. The *subjective_reduction_ratio* of the test set was 79.83%, indicating that the test set could screen out 79.83% of the similar images per the subjective evaluation principle of similarity, with the abnormal recall being 100%.

The method proposed in this study screened out 66.72% of the redundant images without missing the diagnosis of common small intestinal lesions, i.e., venous abnormalities, nodules, lumps, polyps, erythema, spots, white spots, erosions, ulcers, aphthous ulcers, diverticula, granules, redness, abnormal villi, paleness, hookworms, and bleeding. Of the redundant images, 79.09% were screened out without missing venous abnormalities, polyps, erythema, white spots, erosions, aphthous ulcers, diverticula, granules, redness, abnormal villi, paleness, hookworms, and bleeding. The image reduction ratio of the current color structural similarity (CSS) method [6] reaches 93.87% without missing the detection of bleeding lesions. The motion-estimate method [8] can remove 68% of redundant images, although there is no description of lesion retention. The non-negative matrix factorization (NNMF) method [2] achieves an image reduction ratio of 85% with no missed diagnoses of ulcers, venous abnormalities, and white spots. Table 2 is a comparison of the experimental results in this study and those of current methods.

**Table 2. Comparison of experimental results with those of current methods**

| Study / Reduction Score(%) | subjective evaluation | Our work(parameter1*) | Our work(parameter2*) | CSS[6] | Motion-estimate [8] | NNMF[2] |
|---|---|---|---|---|---|---|
| ER-rate | 79.83 | 66.72 | 79.09 | 93.87 | 68 | 85 |
| abnormal-recall | 100 | 100 | 94.23 | / | / | / |
| Protruding lesions-venous structure | 100 | 100 | 100 | / | / | 100 |
| Protruding lesions-Nodule | 100 | 100 | 83.78 | / | / | / |
| Protruding lesions-Mass/tumor | 100 | 100 | 92.86 | / | / | / |
| Protruding lesions-Polyp(s) | 100 | 100 | 100 | / | / | / |
| Flat lesions-Plaque(red) | 100 | 100 | 100 | / | / | / |
| Flat lesions-Spot | 100 | 100 | 100 | / | / | / |
| Flat lesions-Plaque(white) | 100 | 100 | 100 | / | / | 100 |
| Excavated lesion-Erosion | 100 | 100 | 100 | / | / | / |
| Excavated lesion-Ulcer | 100 | 100 | 85.71 | / | / | 100 |
| Excavated lesion-Aphtha | 100 | 100 | 100 | / | / | / |
| Excavated lesion-Diverticulum | 100 | 100 | 100 | / | / | / |
| Mucosa-Granular | 100 | 100 | 100 | / | / | / |
| Mucosa-Erythematous | 100 | 100 | 100 | / | / | / |
| Mucosa-Edematous | 100 | 100 | 100 | / | / | / |
| Mucosa-Pale | 100 | 100 | 100 | / | / | / |
| Content-Parasites | 100 | 100 | 100 | / | / | / |
| Content-Blood | 100 | 100 | 100 | 100 | / | / |

*:parameter1:T1=0.48,T saliency-ssim=0.03;:parameter2:T1=0.88,T saliency-ssim=0.03

With parameter settings based on an abnormal recall of 100%, the proposed similar WCE image screening method based on structural similarity analysis and the hierarchical clustering of visually salient sub-image blocks was introduced into the "redundant image screening" function of the OMOM image workstation produced by Chongqing Jinshan Science & Technology (Group) Co., Ltd. With 4 pictures on 1 screen, the images were played at the speed of 10 frames/second. Figure 9 shows the setting of the play parameters.

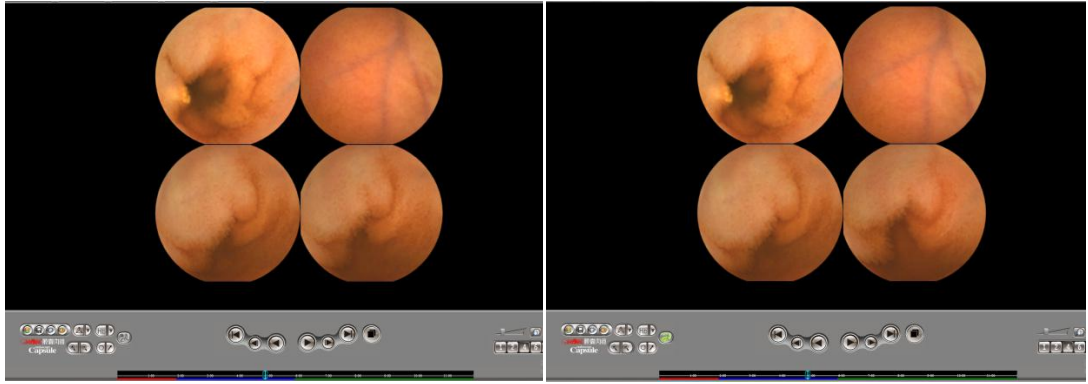

(a) Original images play mode  (b) Redundant images screening play mode

**Figure 9. Setting of play parameters**

Figure 10 presents the original image sequence of the small intestinal capsule endoscopic images. Figure 11 shows the results of the original image sequence without missed redundant image screening.

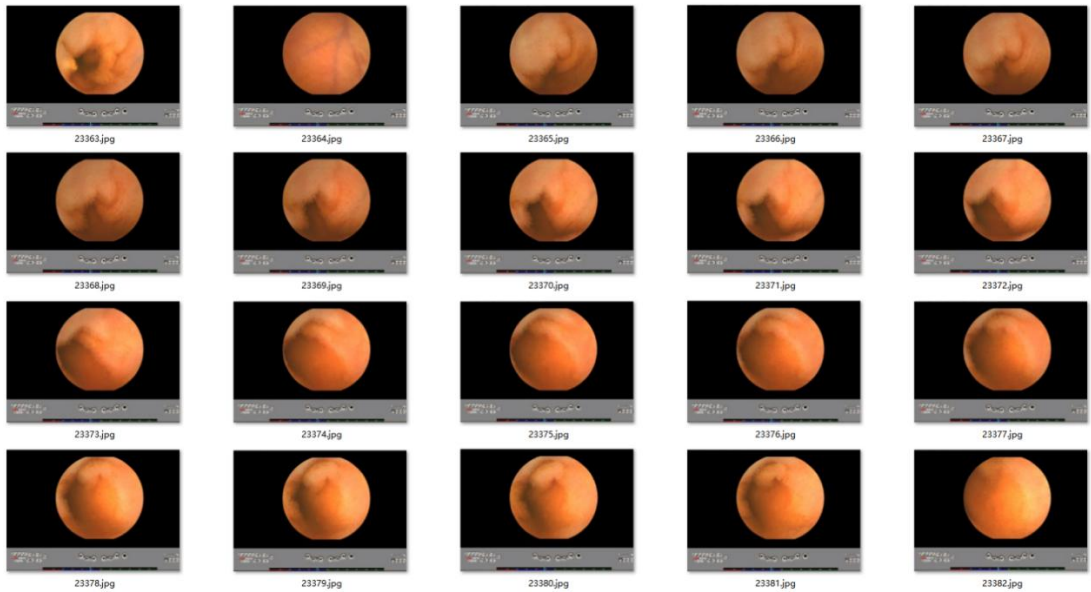

**Figure 10. Original image sequence of the small intestinal capsule endoscopic images in the test set**

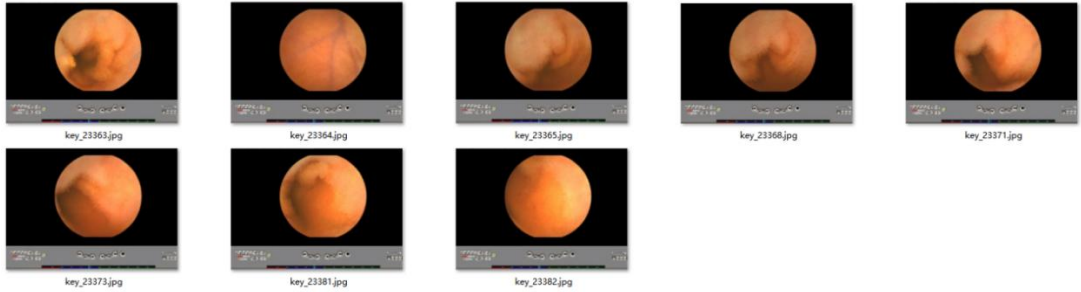

**Figure 11. Results of the original image sequence without missed redundant image screening**

Table 3 lists the average play time and image reduction ratio of all 52 cases in the test set before and after applying the screening of redundant images without missed diagnosis. Before applying the screening of redundant images without missed diagnosis, the average play time was 78 minutes, while after applying the screening, the average play time was reduced to 18 minutes. The similar image reduction ratio of the 52 cases achieved an average of 76%.

**Table 3. Statistics of screened images of 52 cases without missed diagnosis in the test set**

| Patient's No. | Original images | Keyframe images | ER-rate | Original mode play time(minutes) | Redundant image screening mode play time(minutes) |
|---|---|---|---|---|---|
| 1 | 50218 | 15939 | 68.3% | 51.03 | 16.20 |
| 2 | 89824 | 14213 | 84.2% | 91.28 | 14.44 |
| 3 | 76718 | 14692 | 80.8% | 77.96 | 14.93 |
| 4 | 76447 | 16991 | 77.8% | 77.69 | 17.27 |
| 5 | 72031 | 19839 | 72.5% | 73.20 | 20.16 |
| 6 | 61464 | 17289 | 71.9% | 62.46 | 17.57 |
| 7 | 105282 | 22566 | 78.6% | 106.99 | 22.93 |
| 8 | 72209 | 17138 | 76.3% | 73.38 | 17.42 |
| 9 | 17062 | 4006 | 76.5% | 17.34 | 4.07 |
| 10 | 44260 | 8580 | 80.6% | 44.98 | 8.72 |
| 11 | 86851 | 22852 | 73.7% | 88.26 | 23.22 |
| 12 | 105453 | 23856 | 77.4% | 107.17 | 24.24 |
| 13 | 47519 | 11887 | 75.0% | 48.29 | 12.08 |
| 14 | 49678 | 12372 | 75.1% | 50.49 | 12.57 |
| 15 | 35659 | 9186 | 74.2% | 36.24 | 9.34 |
| 16 | 67071 | 22167 | 66.9% | 68.16 | 22.53 |
| 17 | 91231 | 22492 | 75.3% | 92.71 | 22.86 |
| 18 | 95167 | 20958 | 78.0% | 96.71 | 21.30 |
| 19 | 87999 | 16350 | 81.4% | 89.43 | 16.62 |
| 20 | 87941 | 20442 | 76.8% | 89.37 | 20.77 |
| 21 | 92538 | 22071 | 76.1% | 94.04 | 22.43 |
| 22 | 92703 | 17105 | 81.5% | 94.21 | 17.38 |
| 23 | 88319 | 17741 | 79.9% | 89.75 | 18.03 |
| 24 | 85087 | 24819 | 70.8% | 86.47 | 25.22 |
| 25 | 75743 | 21336 | 71.8% | 76.97 | 21.68 |
| 26 | 87839 | 27015 | 69.2% | 89.27 | 27.45 |
| 27 | 90931 | 19256 | 78.8% | 92.41 | 19.57 |
| 28 | 37951 | 6880 | 81.9% | 38.57 | 6.99 |
| 29 | 96128 | 23326 | 75.7% | 97.69 | 23.71 |
| 30 | 63295 | 14870 | 76.5% | 64.32 | 15.11 |
| 31 | 93407 | 23692 | 74.6% | 94.92 | 24.08 |
| 32 | 86495 | 22364 | 74.1% | 87.90 | 22.73 |
| 33 | 87583 | 22798 | 74.0% | 89.01 | 23.17 |
| 34 | 94943 | 21417 | 77.4% | 96.49 | 21.77 |
| 35 | 100385 | 20101 | 80.0% | 102.02 | 20.43 |
| 36 | 38572 | 13965 | 63.8% | 39.20 | 14.19 |
| 37 | 102308 | 29622 | 71.0% | 103.97 | 30.10 |
| 38 | 94106 | 20822 | 77.9% | 95.64 | 21.16 |
| 39 | 72571 | 16845 | 76.8% | 73.75 | 17.12 |
| 40 | 60626 | 16760 | 72.4% | 61.61 | 17.03 |
| 41 | 86207 | 19467 | 77.4% | 87.61 | 19.78 |
| 42 | 89023 | 19549 | 78.0% | 90.47 | 19.87 |
| 43 | 92575 | 21707 | 76.6% | 94.08 | 22.06 |
| 44 | 84511 | 18659 | 77.9% | 85.88 | 18.96 |
| 45 | 112159 | 23852 | 78.7% | 113.98 | 24.24 |
| 46 | 92718 | 18624 | 79.9% | 94.22 | 18.93 |
| 47 | 43592 | 10274 | 76.4% | 44.30 | 10.44 |
| 48 | 49254 | 13307 | 73.0% | 50.05 | 13.52 |
| 49 | 55368 | 12971 | 76.6% | 56.27 | 13.18 |
| 50 | 57086 | 15224 | 73.3% | 58.01 | 15.47 |
| 51 | 57183 | 16968 | 70.3% | 58.11 | 17.24 |
| 52 | 86655 | 13759 | 84.1% | 88.06 | 13.98 |
| Average value | 76307 | 18096 | 76% | 78 | 18 |

T1=0.48/T saliency-ssim=0.03, played at the speed of 10 frames/second, with 4 pictures per screen

## 4. DISCUSSION

First, a test set of WCE images with common small intestinal lesions, i.e., venous abnormalities, nodules, lumps, polyps, erythema, spots, white spots, erosions, ulcers, aphthous ulcers, diverticula, granules, redness, abnormal villi, paleness, hookworms, and bleeding, was established. Unlike most current studies that universally verify a single type of disease, the test dataset in this study can more objectively and comprehensively evaluate the performance of similar image reduction in terms of abnormal recall.

In this investigation, the performance of the proposed algorithm, which is more comprehensive and objective than the currently available evaluation methods, was subjectively and objectively evaluated based on its similarity reduction of a test set. In terms of subjective evaluation, a subjective reduction ratio was calculated based on the evaluation criteria of gastroenterologists. The subjective reduction ratio to some extent represents the maximum reduction ratio that can be reached based on image similarity. The subjective reduction ratio of the test set in this study was 79.83%. Hence, the performance of the proposed algorithm was evaluated with 79.83% as the maximum reduction ratio that could be achieved. When it comes to objective evaluation, with similar image reduction ratio and abnormal recall as the indices, the performance of the algorithm was evaluated under different parameters. In this way, the image reduction ratio that the algorithm could attain without missing the diagnosis of any lesions could be calculated, in addition to the abnormal recall of the algorithm when the subjective reduction ratio is approached. The experimental results revealed that when the abnormal recall in the test set was 100%, the similar image reduction ratio was 66.72%; when the similar image reduction ratio was 79.09%, which was close to the subjective reduction ratio, the abnormal recall was 94.23%. For these results, the missed lesions were usually abnormalities such as dots and single lymphoid follicles that would not affect clinical diagnosis.

The similar WCE image screening method based on structural similarity analysis and the hierarchical clustering of visually salient sub-image blocks proposed in this study was introduced into the "redundant image screening" function of the OMOM image workstation produced by Chongqing Jinshan Science & Technology (Group) Co., Ltd. With retaining an abnormal recall of 100% in the test set as the standard, all small intestinal capsule endoscopic images of the 52 cases in the test set were played at the speed of 10 frames/second, with 4 pictures per screen. Before applying the screening of redundant images without missed diagnosis, the average play time was 78 minutes, while after applying the screening, the average play time was reduced to 18 minutes, and the average image reduction ratio reached 76%, which relieved the doctors' workload when viewing the images without affecting their judgment of lesions.

**Limitations:** Although the abnormal recall of the proposed algorithm is as high as 94.23% when the subjective reduction ratio is approached, and 14 kinds of abnormalities with diagnostic significance are not omitted in the test set, there is still a small risk of missing lesions. Hence, in order to strictly reduce the risk of missing lesions, parameter setting with a smaller reduction ratio is usually adopted in practical application. In the future, the lesions that are missed when the image reduction ratio approaches the screening rate of the human eye should be analyzed in a large-scale test set, in order to design features to better identify those easily missed lesions. Together with the algorithm proposed in this study, the analysis results will further improve the image reduction ratio and reduce the doctors' workload when viewing the images while maintaining the continuity of small intestinal capsule endoscopic images.

## 5. CONCLUSIONS

A method for screening similar WCE images was proposed in this study, based on structural similarity analysis and the hierarchical clustering of visually salient sub-image blocks. Based on clustering, the keyframes in the image dataset were identified using the method of extracting structurally similar keyframes in visually salient sub-image blocks, in order to discern different images more accurately. At an image reduction ratio of 66.72%, the abnormal recall of 17 abnormalities in the test set, i.e., venous abnormalities, nodules, lumps, polyps, erythema, spots, white spots, erosions, ulcers, aphthous ulcers, diverticula, granules, redness, abnormal villi, paleness, hookworms, and bleeding, was 100%. An image reduction ratio of 79.09% is close to the screening rate of the human eye, and the abnormal recall for this amount of reduction was 94.23%. The proposed algorithm was applied to the small intestinal capsule endoscope imaging workstation developed by Chongqing Jinshan Science & Technology (Group) Co., Ltd. After testing 52 complete cases and screening out the redundant images, the average play time was reduced from 78 minutes to 18 minutes, thereby effectively relieving the doctors' workload for viewing images and improving their work efficiency, enabling them to serve more patients, as well as better identify and treat small intestinal lesions.